\documentclass[10pt]{article}
\usepackage{amsmath}
\usepackage{amssymb}
\usepackage[pdftex]{graphicx}
\usepackage{gentium}
\usepackage{microtype}
\usepackage{hyperref}
\DeclareMathSizes{11}{11}{10}{10}  

\newcommand{\defeq}{\mathrel{\mathop:}=}

\newcommand{\mrm}{\mathrm}

\newcommand{\ket}[1]{{\left| {#1} \right\rangle}}
\newcommand{\bra}[1]{{\left\langle {#1} \right|}}

\newcommand{\Evec}{\vec{\mathcal{E}}}

\newcommand{\Esca}{\mathcal{E}}

\begin{document}
\title{Electric dipoles on the Bloch sphere}
\author{Amar C. Vutha \\ {\small Dept. of Physics \& Astronomy, York Univerity, Toronto ON M3J 1P3, Canada} \\ {{\small email: avutha@yorku.ca}}}
\date{ }
\maketitle

\abstract{The time evolution of a two-level quantum mechanical system can be geometrically described using the Bloch sphere. By mapping the Bloch sphere evolution onto the dynamics of oscillating electric dipoles, we provide a physically intuitive link between classical electromagnetism and the electric dipole transitions of atomic \& molecular physics.}

\section{Introduction}
Coherent interactions between electromagnetic fields and quantum mechanical two-level systems are conveniently represented using a geometrical construction known as the Bloch sphere (e.g.\ \cite{Feynman1957,Nielsen2010,Fox2006}). Mathematically, the Bloch sphere is the mapping between the set of two-level density matrices $\rho$ and the set of three-dimensional vectors $\vec{b}$:
\begin{equation}\label{eq:bs}
\rho = \frac{\mathbb{I} + \vec{b} \cdot \vec{\sigma}}{2},   \ \ (|\vec{b}| \leq 1)
\end{equation}
where $\mathbb{I}$ is the $2\times2$ identity matrix, and $\vec{\sigma}$ are the Pauli matrices. The Bloch sphere has become an essential tool to intuitively understand the time evolution of two-level systems in areas ranging from quantum computing \cite{Nielsen2010,Devoret2014} to atomic clocks \cite{Martin2013}. It is also a useful pedagogical tool to introduce concepts such as Rabi oscillations and Ramsey interferometry \cite{Haroche2006}.

Due to the historical sequence of development of atomic and optical physics, from magnetic resonance via lasers to modern atomic physics, the Bloch sphere is usually introduced using magnetic moments and spin precession of spin-1/2 systems in magnetic fields. However, as often as not, the two-level systems encountered in atomic physics involve electric dipole ($E1$) transitions driven by electric fields (e.g.\ the majority of allowed transitions in optical and microwave spectroscopy are of this type). Even though the mathematical construct of the Bloch sphere can be applied to any two-level system (or ``pseudo-spin''), the intuition developed using magnetic dipoles precessing in magnetic fields does not carry over smoothly to electric dipole transitions.

In this article, we demonstrate a connection between the mathematics of Bloch sphere evolution and the dynamics of oscillating electric dipoles. By making this connection, we provide a physically intuitive picture of $E1$ transitions. This framework also evinces the relation between the classical picture of light-matter interactions (``electrons attached to springs'') and the quantum-mechanical description of $E1$ transitions.

\section{Electric dipole transitions}
Consider an atom or molecule, with two states that are connected by an electric dipole ($E1$) transition. Let these states $\ket{0}$ and $\ket{1}$ be separated by an energy $\hbar \omega_0$ in the absence of an external field. The interaction of this system with an electric field $\Evec(t)$ is described by the Hamiltonian $H_\mrm{int} = -\vec{D} \cdot \Evec(t)$, where $\vec{D}$ is the dipole moment operator. Only the off-diagonal matrix elements, $\bra{0}\vec{D}\ket{1}$ and $\bra{1}\vec{D}\ket{0}$, are non-zero. Further, depending on the angular momentum quantum numbers of the states $\ket{0}, \ket{1}$, only one spatial component of the dipole moment vector has a non-zero matrix element. In other words, the system only responds to electric fields with one particular polarization and the problem essentially only involves scalars (this is one of the differences compared to spin-1/2 magnetic moments in a magnetic field). Without loss of generality, set this direction to be $\hat{z}$, let the electric field $\Evec(t) = \Esca(t) \hat{z}$, and let $D \defeq \bra{0}\vec{D}\cdot \hat{z}\ket{1}$ be the matrix element of the dipole moment. The Hamiltonian matrix of the 2-state system in the $\{\ket{0},\ket{1}\}$ basis is
\begin{equation} \label{eq:1}
H = \left( \begin{array}{cc} 0 & -D \Esca(t) \\ -D \Esca(t) & \hbar \omega_0 \end{array} \right).
\end{equation}

The density matrix $\rho$ evolves in time according to the Liouville-von Neumann equation, $i \hbar \frac{d \rho}{dt} = [H,\rho]$. Following the procedure described in \cite{Feynman1957} (or equivalently, using the definition in Equation (\ref{eq:bs})), we rewrite it in terms of the Bloch vector $\vec{b} = (u,v,w)^T$ derived from the density matrix elements,
\begin{equation}\begin{split} \label{eq:uvw}
u & = \rho_{01} + \rho_{10} \\
v & = -i (\rho_{01} - \rho_{10}) \\
w & = \rho_{00} - \rho_{11},
\end{split}\end{equation}
to obtain the Bloch equations for the time evolution of $\vec{b}$.
\begin{equation} \label{eq:bloch}
\frac{d}{dt} \left( \begin{array}{c} u \\ v \\ w \end{array} \right) = \left( \begin{array}{ccc} 0 & \omega_0 & 0 \\ -\omega_0 & 0 & -2\Omega \\ 0 & 2\Omega & 0 \end{array} \right)  \left( \begin{array}{c} u \\ v \\ w \end{array} \right),
\end{equation}
where $\hbar \Omega(t) \defeq D \Esca(t)$. The effects of thermalization and dephasing, usually inserted using the phenomenological relaxation rates $1/T_1, 1/T_2$, have been set aside in order to make the essential point clear. Mathematical solutions of the Bloch equations and their pictorial representations on the Bloch sphere are well known, and can be found in e.g.\ \cite{Feynman1957,Nielsen2010,Fox2006}. We will not repeat them here. Instead, we explore the behaviour of the physical quantities associated with an oscillating dipole.

Consider a non-interacting homogeneous ensemble of two-level systems in a unit volume, e.g.\ a dilute gas of atoms or polar molecules. The Bloch vector's cartesian components $u,v,w$ are related to the following macroscopic quantities:
\begin{eqnarray}
\textrm{Polarization}, P &=& D \ u \label{eq:a}\\
\textrm{Current density}, J &=& \frac{d P}{d t} = \omega_0 D \ v \label{eq:b}\\
\textrm{Internal energy}, U &=&  \hbar \omega_0 \ \frac{(1-w)}{2}. \label{eq:c}
\end{eqnarray}
Equation (\ref{eq:a}) follows from the classical definition of the polarization as the average dipole moment per unit volume. In combination with (\ref{eq:uvw}), it illustrates the important fact that the dipole moment operator $D$ only has a nonzero average $\langle D \rangle$ in a superposition of the basis states $\ket{\psi} = \alpha \ket{0} + \beta \ket{1}, (\alpha,\beta \neq 0)$. Equation (\ref{eq:b}) arises from the definition of the dipole current density $J = \frac{dP}{dt}$ in electromagnetism, and the Bloch equation (\ref{eq:bloch}) for $\frac{du}{dt}$. Equation (\ref{eq:c}) follows from the definition of the density matrix elements $\rho_{00}, \rho_{11}$ as the populations in the two levels, and represents the average energy stored in the ensemble's internal degrees of freedom: $U=\hbar \omega_0$ when every two-level system in the ensemble is in its excited state $\ket{1}$, and $U=0$ when every two-level system is in its ground state $\ket{0}$.

The polarization, current density and internal energy can also be defined as the expectation values of the corresponding operators: 
\begin{equation}\begin{split}
\hat{P} & = D \ (\ket{0}\bra{1} + \ket{1}\bra{0}) \\
\hat{J} & = -i \omega_0 D \ (\ket{0}\bra{1} - \ket{1}\bra{0}) \\
\hat{U} & = \frac{\hbar \omega_0}{2} \left[ \mathbb{I} - \left(\ket{0}\bra{0} - \ket{1}\bra{1}\right) \right].
\end{split}\end{equation}

We can now rewrite the Bloch equations in terms of these physical quantities as
\begin{eqnarray}
\frac{dP}{dt} & = &  J \label{eq:A}\\
\frac{d J}{dt} & = &  -\omega_0^2 P - \frac{2 D^2 \omega_0}{\hbar} \left(1 - \frac{2U}{\hbar \omega_0} \right) \  \Esca(t)  \label{eq:B}\\
\frac{dU}{dt} & = &  - J \Esca(t) \label{eq:C}
\end{eqnarray}

Equation (\ref{eq:A}) is again just the definition of the current density $J$. Combined with the first half of (\ref{eq:B}): $dJ/dt = - \omega_0^2 P$, it describes a simple harmonic oscillator with charge $q$ and mass $m$, whose position $x = P/q$ and momentum $p = m J/q$  oscillate at the natural frequency $\omega_0$. The equatorial plane of the Bloch sphere, spanned by $u$ and $v$, is precisely the phase space plane of this harmonic oscillator.  As the harmonic oscillator oscillates freely, its motion in phase space traces out a circle. Equation (\ref{eq:C}) is just the classical rate at which work is done on the dipole by the electric field. 

\begin{figure}[h!]
\centering
\includegraphics[width=0.75\columnwidth]{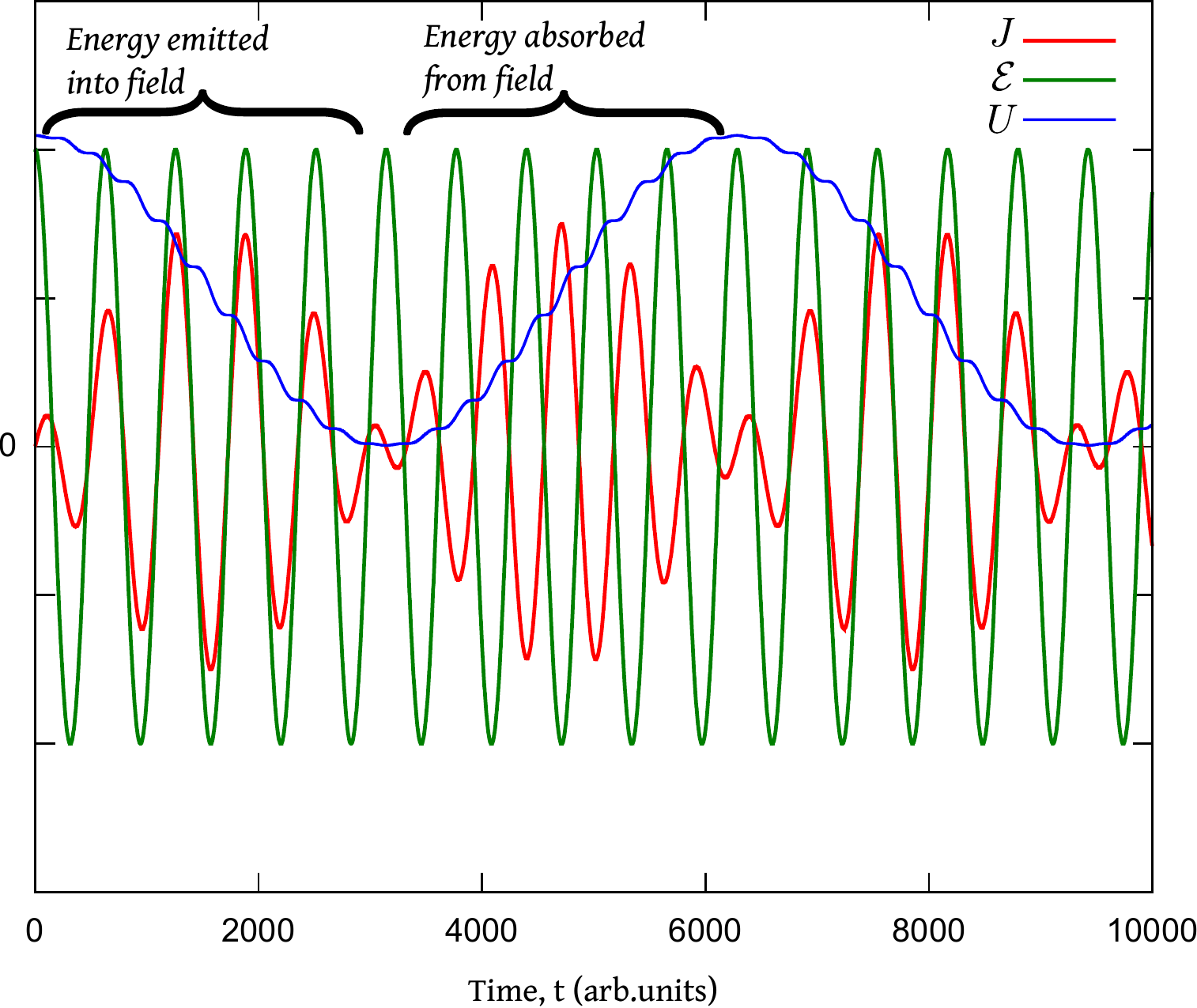}
\caption{\em \small Rabi flopping of an electric dipole. Numerically solving the Schrodinger equation with the Hamiltonian in (\ref{eq:1}), and using the definitions of $P,J,U$ in terms of the density matrix elements, results in this plot. The current density $J$ and internal energy $U$, in response to a resonant ($\omega = \omega_0$) driving electric field $\Esca = \Esca_0 \cos \omega t$, are shown. The current density is in phase with the electric field during the emission part of the cycle, and out of phase during the absorption part of the cycle.} 
\label{fig:rabi_flopping}
\end{figure}

The interesting piece is contained in the second half of (\ref{eq:B}). It describes the oscillator's response to the driving electric field, $\frac{dJ}{dt} = K \Esca(t)$, with a coupling strength $K \defeq - \frac{2 D^2 \omega_0}{\hbar} \left(1 - \frac{2U}{\hbar\omega_0} \right)$. The difference compared to a classical forced oscillator is that \emph{the coupling strength is modulated by the internal energy $U$ of the oscillator}. This is the non-classical nub of the matter (note the $\hbar$s) -- it makes the dynamics nonlinear and leads to the phenomenon of Rabi flopping, illustrated in Figure \ref{fig:rabi_flopping}, and described as follows for a resonant driving field $\Esca(t) = \Esca_0 \cos\omega_0 t$. 

Let the two-level systems in the ensemble initially be in their ground states ($U=0$). Then the coupling constant $K < 0$, so that the current density $J$ is driven 180$^\circ$ out of phase with the electric field $\Esca(t)$ and increases in amplitude. This leads to work being done on the oscillator according to Equation (\ref{eq:C}), and the internal energy $U$ increases. As it increases past the point where the two-level systems have absorbed $\hbar\omega_0/2$ worth of energy on average, the coupling constant $K$ changes sign and the amplitude of the current density begins to be driven back to zero. Eventually, $U \to \hbar \omega_0$, and 1 photon's worth of energy has been absorbed from the electric field. This is the description of a ``$\pi$-pulse'' \cite{Nielsen2010,Fox2006}. The above process then runs in reverse, beginning with $K > 0$, and proceeding until the average internal energy $U \to 0$ again.

In the limit when the rate of work done on the system is negligible,\footnote{This corresponds to a weak absorption/emission of photons, either due to a weak electric field, or because the frequency, $\omega$, of the electric field is very different from the resonance frequency $\omega_0$.} $U,K$ are approximately constant and the dynamics resembles that of a classical driven harmonic oscillator. This explains the great utility of modeling dielectrics in classical electromagnetism as simple harmonic oscillators driven by electromagnetic waves.

In summary, the Bloch equations for a two-level system can be mapped onto the equations of motion of a (nonlinear) harmonically oscillating electric dipole responding to a driving electric field. With this framework, we can interpret the dynamics of two-level systems without any reference to spins, precession or (pseudo-)magnetic fields. The evolution of the density matrix (or state vector) of the system has a simple description in terms of macroscopic physical quantities that are familiar from classical electromagnetism.

\bibliography{dipoles}
\end{document}